\documentstyle[12pt]{article}
\newcommand{\eps}{\epsilon}
\newcommand{\al}{\alpha}
\newcommand{\G}{\Gamma}
\newcommand{\be}{\begin{equation}}
\newcommand{\ee}{\end{equation}}
\newcommand{\ba}{\begin{eqnarray}}
\newcommand{\ea}{\end{eqnarray}}

\title{
A Generalization of Haldane state-counting procedure and
$\pi$-deformations of statistics.
}
\author{
A.V.Ilinskaia$^{1}$, $\quad$
K.N. Ilinski $^{2,3}$\thanks{E-mail: kni@th.ph.bham.ac.uk}$\quad$ and
$\quad$ J.M.F. Gunn$ ^{2}$,\thanks{E-mail: jmfg@th.ph.bham.ac.uk} \\
{\small\it $^{1}$ Institute of Physics, Saint Petersburg University,}
\\
{\small\it Universitetskiy prospect 1,
           191904 Saint Petersburg,} \\
	    {\small\it Russian Federation} \\
[0.5cm]
{\small\it $^{2}$ School of Physics and Space Research,
University of Birmingham,}
\\
{\small\it Birmingham B15 2TT, United Kingdom} \\
[0.5cm]
{\small\it $^{3}$ Institute of
Spectroscopy, Russian Academy of Sciences,} \\
{\small\it Troitsk, Moscow region,
142092, Russian Federation}}
\date{  }

\begin{document}
\setcounter{page}{0}
\maketitle
\vskip -9.5cm
\vskip 9.5cm
\thispagestyle{empty}
\begin{abstract}
We consider the generalization of Haldane's state-counting procedure to
describe all possible types of exclusion statistics which are linear in the
deformation parameter $g$. The statistics are parametrized by elements of
the symmetric group of the particles in question.  For several specific
cases we determine the form of the distribution
functions which generalizes results obtained by Wu. Using them we
analyze the low-temperature behavior and thermodynamic properties of these
systems and compare our results with previous studies of the thermodynamics
of a gas of $g$-ons. Various possible physical applications of these
constructions are discussed.
\end{abstract}

\vspace{1cm}
hep-th/9506111
\newpage

\section{Introduction}
 The importance of the notion of statistics for
the formulation of the quantum many-body problem and investigation of its
macroscopic properties  is very well established. Moreover in the last
decade it was realized that there are dynamical models where
interparticle forces can be regarded as purely statistical
interactions. The most famous of these is a system of particles with
Aharonov- Bohm type interactions \cite{W1,L} (anyons in 2+1 dimensions) and
models solved by the Thermodynamic Bethe Anzatz~\cite{H,WB,Ha}. These two
examples demonstrate the possibility of constructing intermediate
statistics which are neither Bose-Einstein nor Fermi-Dirac statistics.

It is obvious that there are at least two ways to deform statistics to
interpolate between Fermi and Bose statistics. The first is to
deform the exchange factor which appears when particles are permuted
(exchange statistics) and the second is to change the allowed occupation
numbers for each quantum state (exclusion statistics).  This latter
possibility was initially explored by Haldane~\cite{Hald}.

Here we will summarize the definition of exclusion statistics for one
species of particles, for simplicity.
Following Ref.\cite{Hald} let us consider the system confined to a
finite region for which the number $K$ of independent single particle
states is finite and extensive, i.e. proportional to the size of the region
where the particle resides. Then the statistics are determined by the
consequence of adding a second particle, keeping all coordinates of
the existing particles and external properties (size etc.) of the system
fixed.  In general an $N$-particle wave function with fixed coordinates of
$N-1$ particles can be expanded in a basis of wave functions of the $N$th
particle. It is important that in the presence of $N-1$ particles the
number of allowed single particle states $d(N)$ is not in general equal to
$K$ but can depend on the number $N-1$. If we impose a state homogeneity
condition (i.e. independence of $d(N)$ on the particular choice of state
for the $N-1$ particles) and particle homogeneity condition (i.e.
independence $\frac{d(N+m) - d(N)}{m}$ on $m$) then Haldane's exclusion
statistical parameter $g$ is defined by

\be
g = - \frac{d(N+m) - d(N)}{m}
\label{g}
\ee
for any choice of $N$ and $m$.
Using the homogeneity properties it is possible use the
definition in the form $g=d(1) - d(2)$ from which it immediately follows
that such statistical interactions make their first contribution at the
level of the second virial coefficient.

Applying (\ref{g}) to the usual Bose and Fermi ideal gases gives
$g=0$ for the Bose case (i.e. the number $d(N)$ does not depend on
$N$) and $g=1$ for the Fermi case (i.e. after the inclusion of $N-1$
particles the next particle can occupy only $K-N+1$ states and hence
$d(N)=K-N+1$). These straightforward examples make the definition of
the statistical deformation~(\ref{g}) very attractive.

However one complication arises: the definition of a fractional dimension
for the Hilbert space associated with both single and many particle
states.  This question is also closely related to the  construction of
quantum statistical mechanics for a  system with such `$g$-particles' --
state-counting which is  needed to calculate the entropy  and
other thermodynamical quantities of the system. In his original paper
\cite{Hald} Haldane suggested calculating the dimension of the full Hilbert
space for N-particle states $W$ using (we will call this
expression the Haldane-Wu state-counting procedure)

\be
W = \frac{(d(N)+N-1)!}{(d(N)-1)!  (N)!} \qquad d(N)=K-g(N-1) ,
\label{W}
\ee
which  was subsequently used by many authors \cite{WB,Wu,W2,Ou,other} to
describe the thermodynamic properties of $g$-ons. In our previous paper
\cite{IG} we argued that this state-counting procedure is inconsistent with
a step by step application the original definition of Haldane's
statistics~(\ref{g}). In that paper we showed that  expression~(\ref{W})
is actually one of the two simplest state-counting deformations of the
exclusion statistics and so it is perhaps not surprising that this
possibility was realized in many models.  However the procedure for
state-counting is closely connected with the construction of a fractional
dimensional Hilbert space and could not be performed self-consistently
without it.

In reference~\cite{IG} we tried to resolve at least  some of the
difficulties connected with the fractionality of the Hilbert space
dimension.  Our main suggestion was to treat the notion of Haldane's
dimension, and the corresponding statistics, in a probabilistic spirit.
Motivated by the experience of dimensional regularization we defined the
dimension of a space as a trace of  diagonal `unit operator' where the
diagonal matrix elements are not unity in general but are the probabilities
to find a system in a given state.  These probabilities are then  uniquely
defined by the statistics of the particles with homogeneity properties or,
equivalently, by a statistical interaction.

In this paper we want to consider and classify the generalizations of
the Haldane-Wu state-counting procedure. They  describe all the possible
types of exclusion statistics which depend linearly on the deformation
parameter $g$.  As we will show, these statistics (which we term
$\pi$-statistics) are parametrized by elements of the symmetric group
assorted with the particles in question.  For several particular cases of
deformations we derive the equations for distribution functions which
generalize the equation obtained by Wu. These are then used to analyze
the low-temperature behavior and thermodynamic properties of the systems
and compare our results with the thermodynamics of a gas of $g$-ons studied
earlier.

The paper is organized as follows.
In the next section we describe in more details state-counting procedures
for bosons, fermions and the Haldane-Wu ideal gas. These are used
to motivate the introduction of a general construction of exclusion
statistical deformations defined in section~3. In section~4 we
briefly summarize the main points of the derivation of the thermodynamics
for a gas obeying the Haldane-Wu state-counting procedure. Then
in section 5 we investigate  the simplest example
of $\pi$-statistics associated with the identical permutation. More
sophisticated permutations will be explored in section 6.

\section{State-counting procedure. Simple examples}

Let us consider the familiar counting procedure for many-body states for
fermions and bosons in some detail. It is well-known that the number of
quantum states for $N$ identical fermions occupying a set of
single-particle $K$ states is given by the following expression:

\be
W_f = \frac{K!}{N!(K-N)!} = K\cdot(K-1)\cdot(K-2)\ldots(K-N+1)\cdot
\frac{1}{N!}\ .
\label{2}
\ee
To understand the procedure of the construction we can consider the factor
$K$ as the number of ways to place a first particle in the
system, $K-1$ as the corresponding number for a second particle and so on.
In this procedure we consider the particles as distinguishable and at the
end take into account indistinguishability by the factor $1/N!$.

In a similar manner we interpret the bosonic expression for the
number of possible $N$-particle states, which is

\be
W_b = \frac{(K+N-1)!}{N!(K-1)!} = K\cdot(K+1)\cdot(K+2)\ldots(K+N-1)\cdot
\frac{1}{N!}
\label{1}
\ee
as a product of the number of
ways to add  the $i$-th particle to the system. It is easy to see
that in contrast to the fermionic case, where the addition of each particle
reduces the number of accessible states, in the bosonic case the number
of states {it increases} by  unity when a particle is added. A picture
which  provides an  explanation of this strange fact appeals to  quantum
mechanical arguments.

Initially  we assume that the particles
are distinguishable and so we may associate different creation operators
with them.  These do not commute in general and this is the main
origin of the distinction with particles obeying Boltzmann statistics.
For the first particle noncommutativity  does not contribute and we may
place it in $K$ possible states.  To see why the number of states for the
second particle increases we presume that the Hilbert space of the system
is factorized into a tensor product of the Hilbert spaces for each single
particle state.  Then we obtain $K-1$ ways to place the particle
in the empty states and {\it two} possibilities  for the state
occupied by the first particle:  due to the noncommutativity of the
creation operators for the first ($a_{1}^{+}$) and second ($a_{2}^{+}$)
particles the state vectors  $a_{1}^{+}a_{2}^{+}|0>$ and
$a_{2}^{+}a_{1}^{+}|0>$ are in principle different (in the case of
Boltzmann particles these states are identical).  As a result we obtain
$K-1+2$ accessible states for the second particle. By close analogy with
the previous step, at the third step we have $K-2$ possibilities to add to
empty states and $2\cdot 2$ possibilities to add to the two single-occupied
states. And so on.  As a consequence we obtain the product of factors in
equation~(\ref{1}).  The additional factor of $1/N!$
makes our particles indistinguishable. It is not
difficult to see that for the same reasons for Boltzmann particles we
obtain $\frac{K^{N}}{N!}$ which also can be considered from state-counting
procedure point of view.

So for both fermions and bosons the  expression for $W$ contains a
product of $N$ brackets which are interpreted as the number of the
accessible states for the corresponding particle. We will now  construct in
the same manner intermediate statistics which reduce to fermionic or
bosonic statistics for particular values of the deformational parameter.

We can now state the simplest deformation of the state-counting
procedure, which is natural and interpolates between Bose-
and Fermi- cases. We suppose that the increase in the number of available
single particle states due to the addition of one particle is not  $1$, as
it was for bosons, nor  $-1$, as it was for fermions, but is $\al$. Then
the number for states of $N$ particles occupying a set of $K$ single
particle states is given by

\be
W = \frac{1}{N!}\cdot K \cdot (K+\al) \cdot (K+2\al) \ldots (K+\al(N-1))\ .
\label{5}
\ee
This expression reduces to $W_b$ when $\al=1$ and to $W_f$ when $\al=-1$
in such a manner that the $l$th bracket of Eq.(\ref{1}) is transformed into
the $l$th one of Eq.(\ref{2}). In the next section we will consider this
simple example in detail.

The Haldane-Wu state-counting procedure can also be viewed in this manner.
It gives the following dimension for the  N-particle space:

\be
W = \frac{\bigl(K+(N-1)(1-g)\bigr)!}{N!\bigl(K-gN-(1-g)\bigr)!} \ ,
\label{3}
\ee
which also interpolates between the Bose- and Fermi- expressions.
Originally the expression was
introduced by Haldane~\cite{Hald} to describe a state-counting procedure
for  particles with exclusion statistics
and subsequently was used by many authors
\cite{Wu,WB,W2} to investigate the thermodynamic properties of a  gas of
such particles (g-ons).  One can rewrite (\ref{3}) as follows

\be
W  =  \frac{1}{N!}\cdot
\prod_{l=0}^{N-1}\bigl(K-g(N-1)+l\bigr) \ .
\label{4}
\ee
In this form
we can see again  that  each of the brackets in the expression
interpolates from one of brackets in Eq.(\ref{1}) (by $g=0$) to another one
in Eq.(\ref{2}) (by $g=1$).  Moreover the $l$th bracket of Eq.(\ref{1}) is
transformed into $(n-l+1)$th one of Eq.(\ref{2}), i.e. brackets of
Eq.(\ref{1}) correspond to brackets of Eq.(\ref{2}) in reverse order by
replacement of $g=0$ by $g=1$ in Eq.(\ref{4}).

Motivated by  these two examples, in the next section we give the general
definition of  deformed state-counting procedures which are
based on the linearity of the deformation of single-particle dimensions
and on the symmetric group of the system of particles.

\section{State-counting procedure: general construction}

In this section
we will describe the general case of
state-counting deformations of statistics which are linear in $g$. These
deformations differ from Haldane's original exclusion statistics but
are still of an exclusion type as they are concerned with
occupation number considerations (as in
eqs.(\ref{4}) and (5)).  Moreover we will show that the expressions
(\ref{4}) and (\ref{5}) represent the two simplest examples of such
deformations.

The definition of the deformed statistics (which depend linearly on $g$)
consists in stating a one to one correspondence between the brackets of
Eq.(\ref{1}) and  Eq.(\ref{2}).
It is clear that there are $N!$ possibilities of such
a correspondence and they are parametrized by the general element of the
group of permutations of the set  of $N$ distinguishable
objects. The discussion in the previous section implies that the general
expression for the number of states is:

\be
W  =  \frac{1}{N!}\cdot
\prod_{l=0}^{N-1}\bigl(K-gl+(1-g)\pi_l\bigr) \ ,
\label{6}
\ee
where $\pi_l$ is the $l$th member of a permutation of $1,\ldots ,N$,
denoted by $\pi$.  The cases $g=0$ and $g=1$ correspond respectively to
bosons and fermions independently of the choice of $\pi$.  For the
identical permutation ($\pi =id:\pi_l=l$) expression~(\ref{6}) reduces to
Eq.(\ref{5}) with $\al = 1-2g$ while for the inverse case ($\pi_l = N-l-1$)
we find Eq.(\ref{4}).  It is obvious that  expression (\ref{6})
represents all possible linear (in term of $g$) exclusion statistics.
Henceforth we will term the statistics parametrized by the permutation
$\pi$ as {\it $\pi$-statistics}.

Let us now note that not all permutations $\pi$ lead to distinct
thermodynamic behavior in the thermodynamic limit $K,N\rightarrow
\infty$, $N/K=const$. Indeed, permutations differing only by a finite
number of pairwise transpositions give the same statistics in the
thermodynamic limit. Moreover only the permutations which differ by an
infinite number, $M=O(N)$, of pairwise transpositions which in their
order change the positions of the particles on infinite number
$M^{\prime}=O(N)$ give distinguishable thermodynamic results. So
if we are interested in  thermodynamic quantities we
should consider not the original group of  permutations $S$ but the
quotient group $S^{\prime}=S/H$ where $H$ is the subgroup of permutations
for which the  numbers $M$ and $M^{\prime}$ are $o(N)$ (we imply that an
arbitrary but finite number of elements in the products are allowed).
The simplest example of such transposition $\pi \in H$ can be easy found:
$\pi_{l}=l+1$ for $l=0,\ldots,N-2$ and $\pi_{N-1}=0$.  In this case
$M^{\prime}$ is equal $N$ but $M=1$ and is negligible if $N\rightarrow
\infty$. In the following we will be interested specifically in
$\pi$-statistics with $\pi \in S^{\prime}$ for $N\rightarrow \infty$.

Following \cite{Wu} we can consider systems containing particles of
different species. We introduce the notation: $K_i$ is the number of
independent states of a single particle of species $i$,
$N_i$ is the number of particles of species $i$. Then the number of
many-body states at fixed $\{N_i\}$ in the framework of
$\pi$-statistics is given by the expression:

\be
W  = \prod_i \frac{1}{(N_i)!}\cdot
\prod_{l=0}^{N_i-1}\bigl(K_i-g_{i}l_i+(1-g_{i})\pi^{(i)}_{l_i}
-g_{ik}\pi^{ik}_{l_{k}}
\bigr) \ ,
\label{7}
\ee
where, as above, $\pi^{i}$ is a permutation of the particles of $i$-th
species and the mapping $\pi^{ik}$ is a mapping of the $k$-species into the
$i$-th one.
If the parameters $g_{ik}$ are not zero or the mappings $\pi^{ik}$
are not trivial ($\pi^{ik}_{l_{k}}=0$)
we will call such $\pi$-statistics as {\it mutual
$\pi$-statistics}. This construction exactly generalizes the Haldane-Wu
mutual statistics  and models the dependence of the number of states for
particles of species $i$ on the particle number $N_j$ of other species.
We will not deal with mutual statistics in detail but the
generalizations are straightforward.

\section{Thermodynamical properties of Haldane-Wu gas}

This section is included to make the picture more complete and the
references more convenient. It is devoted to the brief description of the
thermodynamic properties of a gas using the Haldane-Wu state-counting
procedure (\ref{3}):
$$
W =
\prod_{i}\frac{\bigl(K_{i}+(N_{i}-1)(1-g)\bigr)!}{N_{i}!\bigl(K_{i}-
gN_{i}-(1-g)\bigr)!} \ .
$$
Starting with this counting procedure it is possible to construct
thermodynamics in the standard manner. In the thermodynamic limit, the
number of particles $N_{i}$, as well as the number of single-particle
states $K_{i}$, becomes infinite. But the occupation numbers
$n_{i}=N_{i}/K_{i}$ still remain finite. The entropy of the system
$\mbox{\cal S}=\ln W$ (we set Boltzmann's constant equal to unity). By
definition, the ideal gas with $\{N_{i}\}$ particles has a total energy
of the following form:
$$
E=\sum_{i} N_{i} \epsilon_{i}
$$
with constant
$\epsilon_{i}$. For such gases, the thermodynamic potential $\Omega $ can
be evaluated by the minimizing $$ \Omega = E -T \mbox{\cal S} -\mu \sum_{i}
N_{i} $$ with respect to the variation of the densities $n_{i}$ (here $T$
is the temperature and $\mu$ is the chemical potential).

The resulting thermodynamics may be summarized as follows \cite{Wu}:
$$
\Omega = -T \sum_{i} K_{i} \ln \frac{1+w_{i}}{w_{i}}  \ ,
$$
where the function $w_{i}$ is defined by the following equation:
\be
w_{i}^{g} (1+w_{i})^{1-g} = e^{(\epsilon_{i}-\mu)/T} \ .
\label{WE}
\ee
Using the same notation the distribution functions $n_{i}$ may be expressed
as
\be
n_{i}= 1/(w_{i} + g) \ .
\label{WD}
\ee
Equations (\ref{WE}, \ref{WD}) were initially obtained in~\cite{Ou}
for a system of anyons in a strong magnetic field on the lowest Landau
level.  They lead to fermionic-like behavior for cases with the value of
the deformation parameter $0< g \leq 1$ (i.e. except the case of real
bosons) at low temperatures. In particular, at  zero temperature the
distribution function contains a `Fermi-step':
\be
n_i = \left\{
\begin{array}{ll}
0 & \mbox{ if } \eps_i > E_F  \\
-1/\al & \mbox{ if } \eps_i < E_F
\end{array} \right.   \ .
\label{step}
\ee
We will refer to the formulae (\ref{WE},\ref{WD},\ref{step}) when other
deformed state-counting procedures are considered in the sections below.

\section{id-Statistics and its thermodynamics}

In this section we derive the occupation number distribution for gas of
particles obeying $id$-statistics and compare it with the
result obtained in~\cite{Wu} for the permutation $\pi_{X}=\pi_l = N-l-1$,
i.e. with Haldane-Wu state-counting procedure.

Let us consider the identical permutation in Eq.(\ref{7}):
$\pi_l^{(i)}=l$ for all $i$. In this case we regain expression
(\ref{5}) for the number of states of particles of species $i$ with $\al =
1-2g$. One can rewrite Eq.(\ref{7}) using $\G$-functions (due
to the well-known property of $\G$-function: $\G(z+1)=z\G(z)$) as
follows

$$
W  = \prod_i \frac{\G\left(\frac{K_i+\al N_i}{\al}\right) \al^{N_i}}
{\G\left(\frac{K_i}{\al}\right)(N_i)!}  \ .
$$
We consider an ideal gas of such particles, where the total energy is a
direct sum

$$
E = \sum_i N_i \eps_i  \ ,
$$
where $\eps_i$ is the energy of a particle of species $i$. Following the
standard procedure~\cite{Landau}, one can consider a grand canonical
ensemble at temperature $T$ and with chemical potential $\mu$. Then the
partition function is

$$
Z = \sum_{\{N_i\}} W(\{N_i\}) \exp\left\{\sum_i N_i(\mu-\eps_i)/kT\right\}
\ .
$$
For very large $K_i$ and $N_i$ the summand has a very sharp peak around the
set of most probable particle numbers $\{N_i\}$. Using the
asymptotic approximation for $\G$-functions ($\ln \G(z)=z\ln z$) and
 introducing the average occupation number $n_i=N_i/K_i$, one can express
 $\ln W$ as follows

$$
\ln W = \sum_i K_i \left\{\left( \frac{1}{\al} + n_i\right)
\ln (1+ \al n_i) - n_i \ln n_i \right\} \ .
$$
The most probable distribution of $n_i$ is determined by

$$
\frac{\partial}{\partial n_i} \sum_i K_i \left\{\left( \frac{1}{\al} +
n_i\right) \ln (1+ \al n_i) - n_i \ln n_i
+ n_i \frac{\mu - \eps_i}{kT}\right\} = 0  \ ,
$$
that leads to the expression for the average occupation number:

\be
n_i = \frac{1}{\exp\left(\frac{\eps_i-\mu}{kT}\right) - \al} \ .
\label{12}
\ee
This expression recovers the Bose, Boltzmann and Fermi distributions with
$\al = 1$, $\al=0$ and $\al=-1$ respectively.

Now we consider in detail the thermodynamics of a gas with
$0<\al\leq 1$ and demonstrate Bose-condensation at low temperature
and values of the deformation parameter $\al \geq 0$ (which is
equivalent to the inequality $g \leq 1/2$).
The energy distribution of particles with
the occupation number~(\ref{12}) is~\cite{Landau}

\be
dN_{\eps} = \frac{SVm^{3/2}}{\sqrt{2}\pi^2 h^3} \cdot
\frac{\sqrt{\eps} \  d\eps}{\exp\frac{(\eps-\mu)}{T}- \al} \ ,
\label{13}
\ee
where we use units such that Boltzmann's constant $k=1$, spin degeneracy
factor $S=2s+1$ ($s$ being the spin of the particle), $m$ is the mass of
the particle, $V$ is the total volume of the gas.  Integrating with respect
to $\eps$ we obtain the number of particles with energy $\eps>0$ in the
gas:

\be
\frac{N}{V} = \frac{S(mT)^{3/2}}{\sqrt{2}\pi^2 h^3} \int_0^{\infty}
\frac{\sqrt{z} \  dz}{e^{z-\mu/T} - \al} \ .
\label{14}
\ee
This formula determines the chemical potential $\mu$ of the gas as a
function of its temperature $T$ and density $N/V$.

{}From  the condition $n_i\geq 0$ one can derive a restriction on
$\mu$:

\be
\mu \leq - \ln\al \cdot T\ ,\mbox{ where }\ \ln\al < 0 \ .
\label{15}
\ee
If the temperature of the gas is lowered at constant density $N/V$, the
chemical potential $\mu$ given by~(\ref{14}) increases. It reaches the
limiting value determined by the relation~(\ref{15}) at a temperature
$T_0$, which can be determined from the equation

$$
\frac{N}{V} = \frac{S(mT_0)^{3/2}}{\sqrt{2}\pi^2 h^3 \al} \int_0^{\infty}
\frac{\sqrt{z} \  dz}{e^{z} - 1} \ ,
$$
which gives the following value for $T_0$:

\be
T_0 = \frac{3.31 h^2}{mS^{2/3}}\left( \frac{N}{V} \right)^{\frac{2}{3}}
\cdot \al^{\frac{2}{3}} = T_0^{(b)}\cdot \al^{\frac{2}{3}}
\label{17}
\ee
with the temperature of Bose-Einstein condensation for
bosons ($\al = 1$) $T_0^{(b)}$.

For $T<T_0$ Eq.~(\ref{14}) has no solution obeying the relation~(\ref{15}).
This contradiction arises because we have actually neglected in~(\ref{13})
the particles with $\eps = 0$ by multiplying the expression by
$\sqrt{\eps}$.  In reality the situation for $T<T_0$ is as follows.
Particles with energy $\eps>0$ are distributed according to
formula~(\ref{13}) with $\mu = -\ln\al\cdot T$:

$$
dN_{\eps} = \frac{1}{\al} \cdot \frac{SVm^{3/2}}{\sqrt{2}\pi^2 h^3} \cdot
\frac{\sqrt{\eps} \  d\eps}{e^{\eps/T}- 1} \ .
$$
The total number of particles with energies $\eps>0$ is

$$
N_{\eps>0}= N(T/T_0)^{\frac{3}{2}}
$$
and the number of particles in the lowest state with $\eps=0$ is

\be
N_{\eps=0}= N \left(1 - \left(\frac{T}{T_0}\right)^{\frac{3}{2}}\right)\ .
\label{20}
\ee
Thus we obtain the Bose-Einstein condensation for particles
obeying~(\ref{12}) for  $0<\al\leq 1$. Moreover the form of the
expression~(\ref{20}) does not depend on $\al$; for different values $\al$
we just obtain different values of Bose-Einstein condensation
temperatures~(\ref{17}).

For $-1\leq\al<0$, it follows from~(\ref{12}) that
$$
n_i\leq -\frac{1}{\al} \ .
$$
In this case at $T=0$ the average occupation number for single particle
states with a continuous energy spectrum has a `fermionic' step-like
distribution:

$$
n_i = \left\{
\begin{array}{ll}
0 & \mbox{ if } \eps_i > E_F  \\
-1/\al & \mbox{ if } \eps_i < E_F
\end{array} \right.   \ .
$$

Now we can compare the results of this section with those derived for the
inverse permutation $\pi_{X}$ in~\cite{Wu}.
Let us remember that for the inverse permutation the average occupation
number at $T=0$ has a fermi-like distribution for any value
of the parameter except that which corresponds to bosons. In the present
case for the identical permutation we obtained a fermion-like distribution
for parameter values interpolating between Fermi and Boltzmann
distributions ($0<\al\leq 1$). For parameter  values  between Boltzmann and
Bose distributions ($-1\leq \al <0$) we obtained a boson-like behavior at
low temperatures~(\ref{20}).  Furthermore, for the identical permutation we
have the Boltzmann distribution at $\al=1/2$ while for the inverse one the
distribution tends to the Boltzmann distribution (at the same value of the
parameter) only in the high temperature limit.

Now we are going to consider several more complicated examples of
permutations and compare them with those discussed above.

\section{Other permutations}
As it is easy to see from  Eqn.~(\ref{7}), we have a lot of
possibilities to construct statistics using different permutations. A more
natural and simple way to do this is to divide the set
of brackets into two equal parts (we assume the number of brackets is
even, which is not important in the thermodynamic limit) and then
consider the two simplest permutations (id- and inverse permutations) for
these parts (see Figs.2-5). These constructions seem to be simple,
however  they display a great
variety of thermodynamic properties.

\subsection{The permutation $\pi_{XX}$}
At first we will deal with the permutation where the first $N/2$
factors of the number of states for fermions correspond to those for
bosons crosswise and the same correspondence takes place for the second
$N/2$ brackets~(Fig.2). It is natural to call this permutation
$\pi_{XX}$.

We can formalize our rule representing the permutation in terms of the
following expressions:

$$
{\pi_{XX}}_l^{(1)} = \frac{N}{2} - l - 1 \ ,
\  \mbox{ for } \  0\leq l \leq \frac{N}{2}-1
$$
$$
{\pi_{XX}}_l^{(2)} = \frac{3}{2} N - l - 1 \ ,
\  \mbox{ for } \  \frac{N}{2}\leq l \leq N-1 \ .
$$
Then using Eqn.~(\ref{7})  the number of
many-body states for very large $K_i$ and $N_i$  is given by the equality:

$$
W_{XX}  = \prod_i \frac{1}{(N_i)!}\cdot
\frac{\left(K_i+ \frac{1-g}{2}N_i\right)! }
{\left(K_i- \frac{g}{2}N_i\right)! }
\cdot \frac{\left(K_i+ \left(1 -\frac{3}{2}g\right)N_i\right)! }
{\left(K_i+ \frac{1-3g}{2}N_i\right)! }  \ .
$$
Following the same procedure as in the previous section one can derive the
expression for the average occupation number

$$
{n_{XX}}_i = \frac{1}{w_{XX}(\xi) + g/2}    \ ,
$$
where $\xi = \exp\left(\frac{\eps_i-\mu}{kT}\right)$ and $w_{XX}(\xi)$
satisfies the following equation:

\be
w_{XX}^{\frac{g}{2}}(\xi)\cdot
\left( w_{XX}(\xi) +\frac{1}{2}  \right)^{\frac{1-g}{2}}
\cdot \Bigl(w_{XX}(\xi)+1-g\Bigr)^{1-\frac{3}{2}g} \cdot
\left(w_{XX}(\xi) + \frac{1}{2} - g \right)^{\frac{-1+3g}{2}} = \xi \ .
\label{wXX}
\ee
This equation may be considered as an  analogue of equation
(\ref{WE}). It is interesting to note that not only the general form of
this equation is similar to the form of Eqn.(\ref{WE}): the sum of the
powers in the expression (\ref{wXX}) is equal to unity as in
(\ref{WE}).  We will see that this property is common to the
other permutations.

The average occupation number at $T=0$ has
a fermionic step-like distribution  for all values of  the parameter $g$
except $g=0$ (bosons):

\be
{n_{XX}}_i = \left\{
\begin{array}{lll}
0 & & \mbox{ if } \eps_i > E_F  \\
\frac{2}{g} & \mbox{ for } g\leq 1/2 &  \\
\frac{2}{3g-1} & \mbox{ for } g > 1/2 & \mbox{ if } \eps_i < E_F
\end{array} \right. \ ,
\label{FermiXX}
\ee
where the Fermi level  is a continuous function of $g$ for $0<g\leq 1$
but not a smooth function as it was for the inverse permutation $\pi_{X}$
($E_F=1/g$).
For the case $g=1/2$ it is possible to solve the Eqn.(\ref{wXX}) and
obtain the following expression for the average occupation number:

\be
\left( n_{XX} =
\frac{1}{\sqrt{\frac{1}{16} + \exp(2\frac{\eps-\mu}{kT})}}
\right)_{g=\frac{1}{2}} \ .
\label{1/2XX}
\ee
Comparing this with previous cases
we recall that
for the inverse permutation $\pi_{X}$ the same
expression occurs with the replacement of $1/16$ by $1/4$ ~\cite{Wu} and
in the case of the identical permutation for $g=1/2$ we
have just the Boltzmann distribution.
So in some sense the function $n_{XX}$ is closer to the Boltzmann
distribution at $g=1/2$ than to Wu's distribution.

\subsection{The permutation $\pi_{XI}$}

The second example of this section is generated by the permutation where
the first $N/2$ brackets of the number of states for fermions
correspond to those for bosons crosswise and for the second $N/2$ brackets
the identical permutation takes place~(Fig.3). One can  write the
following expressions for it:

$$
{\pi_{XI}}_l^{(1)} = \frac{N}{2} - l - 1 \ ,
\  \mbox{ for } \  0\leq l \leq \frac{N}{2}-1
$$
$$
{\pi_{XI}}_l^{(2)} =  l  \ ,
\  \mbox{ for } \  \frac{N}{2}\leq l \leq N-1 \ .
$$
Then the number of many-body states  can be expressed as:

$$
W_{XI}  = \prod_i \frac{1}{(N_i)!}\cdot
\frac{\left(K_i+ \left(\frac{N_i}{2}-1\right) (1-g)\right)! }
{\left(K_i- g\left(\frac{N_i}{2}-1\right) -1\right)! }
\cdot \frac{\Gamma\left(
\frac{K_i+ (1-2g)N_i}{1-2g} \right)}
{\Gamma\left(\frac{K_i+ (1-2g)N_i/2}{1-2g}\right) }
\cdot (1-2g)^{\frac{N_i}{2}}  \ .
$$
Following  the same procedure as in previous sections (i.e.minimizing
the thermodynamic potential $\Omega$), we obtain
an expression for the average occupation number:

$$
{n_{XI}}_i = \frac{1}{w_{XI}(\xi) + g/2}   \ ,
$$
where the function $w_{XI}(\xi)$ as usual is defined by the
equation:

$$
w_{XI}^{\frac{g}{2}}(\xi)\cdot
\left(w_{XI}(\xi) + \frac{1}{2} \right)^{\frac{1-g}{2}}
\cdot \left(w_{XI}(\xi) + 1-\frac{3}{2}g\right) \cdot
\left( w_{XI}(\xi) + \frac{1-g}{2}  \right)^{-\frac{1}{2}} = \xi  \ .
$$

For this permutation at $T=0$ we also  obtain
a step-like distribution  for any value of  the parameter $g$
except $g=0$ (bosons):

\be
{n_{XI}}_i = \left\{
\begin{array}{lll}
0 & & \mbox{ if } \eps_i > E_F  \\
\frac{2}{g} & \mbox{ for } g\leq 2/3 &  \\
\frac{1}{2g-1} & \mbox{ for } g > 2/3 & \mbox{ if } \eps_i < E_F
\end{array} \right. \ .
\label{FermiXI}
\ee
As for the permutation $\pi_{XX}$  the Fermi level  is a continuous, but
not smooth, function of $g$ for $0<g\leq 1$.

\subsection{The permutation $\pi_{IX}$}

The last example of such a type of permutation can be denoted by
$\pi_{IX}$, illustrated by Fig.4 and described by the formulae:

$$
{\pi_{IX}}_l^{(1)} =  l  \ ,
\  \mbox{ for } \  0\leq l \leq \frac{N}{2}-1
$$
$$
{\pi_{IX}}_l^{(2)} = \frac{3}{2}N - l - 1 \ ,
\  \mbox{ for } \  \frac{N}{2}\leq l \leq N-1  \ .
$$
To escape repetition of standard arguments, we just state
the main results for this case without comment:

$$
W_{IX}  = \prod_i \frac{1}{(N_i)!}\cdot
\frac{\Gamma\left( \frac{K_i+ (1-2g)N_i/2}{1-2g} \right)}
{\Gamma\left(\frac{K_i}{1-2g}\right) }
\cdot (1-2g)^{\frac{N_i}{2}}  \cdot
\frac{\left(K_i+ \left(1 -\frac{3}{2}g\right)N_i\right)! }
{\left(K_i+ \frac{1-3g}{2}N_i\right)! }  \ ,
$$
$$
{n_{IX}}_i = \frac{1}{w_{IX}(\xi) + \frac{3g-1}{2}} \ ,
$$
\be
w_{IX}^{\frac{3g-1}{2}}(\xi)\cdot
\left( w_{IX}(\xi) + \frac{1}{2}  \right)^{1-\frac{3}{2}g} \cdot
\left( w_{IX}(\xi) + \frac{g}{2} \right)^{\frac{1}{2}}  \cdot  = \xi \ .
\label{wIX}
\ee

In contrast with the previous examples,
for this permutation at $T=0$  the fermion distribution takes place
only for values of the parameter $g>1/3$ and the Fermi level is a
continuous and smooth function of $g$:

$$
{n_{IX}}_i = \left\{
\begin{array}{lll}
0 & & \mbox{ if } \eps_i > E_F  \\
\frac{2}{3g-1} & \mbox{ for } g > 1/3 & \mbox{ if } \eps_i < E_F
\end{array} \right.  \ .
$$
To investigate the behavior of particles with such statistics for
$g\leq 1/3$ we will consider the particular example $g=1/3$.
In this case the equation~(\ref{wIX}) can be solved and the average
occupation number is given by

$$
\left( n_{IX} =
\frac{3}{\sqrt{9\xi^2+\frac{1}{4}} - 1}
\right)_{g=\frac{1}{3}} \ .
$$

The condition $n\geq 0$ implies the following inequality for $\mu$

$$
\mu \leq \ln(2\sqrt{3})kT \ .
$$
It is similar to relation~(\ref{15}) which was obtained for the
statistics generated by the identical permutation. Following the discussion
of the previous section one can derive that  Bose-Einstein
condensation takes place in this case  and the temperature for
condensation is determined by the following equation

$$
\frac{N}{V} = \frac{S(mT_0)^{3/2}}{\sqrt{2}\pi^2 h^3 \al} \int_0^{\infty}
\frac{3\sqrt{z} \  dz}{\frac{1}{2}\sqrt{3e^{z} + 1} - 1} \ .
$$
We obtain the Bose-Einstein condensation at $g=1/3$ and it takes place at
$g=0$ (bosons). So one can conclude that the statistics generated by the
permutation $\pi_{IX}$ obey Bose-like behavior at low temperature
for the values of the parameter $0\leq g\leq 1/3$.

For the permutations $\pi_{IX}$ and $\pi_{XI}$ one can obtain the
equivalent expressions for the average occupation number by $g=1/2$:

$$
\left(n_{XI} =
\frac{1}{\sqrt{2}}\exp\left(-2\frac{\eps-\mu}{kT}\right)
\left({\sqrt{1 + 4^5 \exp(4\frac{\eps-\mu}{kT})}} -1 \right)^{\frac{1}{2}}
\right)_{g=\frac{1}{2}} \ .
$$
In some sense this expression is closer to the Boltzmann distribution
than~(\ref{1/2XX}). It can be connected with the presence of  partially
identical pieces in the basic permutation. Indeed, in the case of pure
identical permutation in this limit we had a Boltzmann distribution.
In other words we can try to characterize the thermodynamic properties
in terms of permutation characteristics.
Let us stress, however, that distribution functions of all the above
examples have the Boltzmann limit at enough high temperature.

\subsection{The permutation $\pi_{\bowtie}$}

Let us now turn to the last example, which is a little more
complicated.
Consider the permutation that correlates the first $N/2$ brackets
of the number of states for fermions with the second  $N/2$
for bosons identically and the same correspondence takes place for
the remaining brackets. It is similar to the identical
permutation but the sets of the first and second $N/2$ brackets are
related in a crosswise fashion~(Fig.5).
We will denote  this permutation by the index $\bowtie$ and describe it by:

$$
{\pi_{\bowtie}}_l^{(1)} = \frac{N}{2} + l  \ ,
\  \mbox{ for } \  0\leq l \leq \frac{N}{2}-1
$$
$$
{\pi_{\bowtie}}_l^{(2)} = l-  \frac{N}{2}  \ ,
\  \mbox{ for } \  \frac{N}{2}\leq l \leq N-1  \ .
$$
The main results are

$$
W_{\bowtie}  = \prod_i \frac{1}{(N_i)!}\cdot
\frac{\Gamma\left( \frac{K_i+ N_i(1-3/2\cdot g)}{1-2g} \right)}
{\Gamma\left(\frac{K_i +N_i(1-g)/2}{1-2g}\right) }
\frac{\Gamma\left( \frac{K_i+ N_i(1-3g)/2}{1-2g} \right)}
{\Gamma\left(\frac{K_i -N_i g/2}{1-2g}\right) }
\cdot (1-2g)^{N_i}  \ ,
$$
$$
{n_{\bowtie}}_i = \frac{1}{w_{IX}(\xi) + \frac{g}{2}} \ ,
$$
\be
w_{\bowtie}^{\frac{g}{2}}(\xi)\cdot
\left(w_{\bowtie}(\xi) + \frac{1}{2} \right)^{\frac{g-1}{2}} \cdot
\left( w_{\bowtie}(\xi) + \frac{1}{2}-g \right)^{\frac{1-3g}{2}}  \cdot
\left( w_{\bowtie}(\xi) + 1-g \right)^{1-\frac{3}{2}g}   = \xi^{1-2g}\ .
\label{wbowtie}
\ee

It is interesting to note that in this case at $T=0$ we obtain exactly the
same results as for the permutation $\pi_{XX}$:
$n_{\bowtie}=n_{XX}$~(\ref{FermiXX}), i.e.  we obtain the Fermi-like
distribution at any value of parameter except that which corresponds to
bosons and the Fermi level  is a continuous, but not a smooth, function of
the parameter.

The equation~(\ref{wbowtie}) obeys the identity at  $g=1/2$. So for this
case we have to consider the original expression for the number of
many-body states for this permutation with $g=1/2$.  Following the standard
procedure one can obtain the equation for the average occupation number:

$$
\left( \sqrt{1-\frac{n_{\bowtie}^2}{16}}  =
n_{\bowtie} \exp\left(\frac{1}{1-\frac{n_{\bowtie}^2}{16}}\right)
\cdot \exp\left(\frac{\eps-\mu}{kT}\right)\right)_{g=\frac{1}{2}} \ .
$$

\subsection{Summary}

We have considered four non-trivial examples of permutations obeying four
different statistics. Summarizing our discussion, we can note that in only
one case, $\pi_{IX}$, we obtained a Bose-like behavior at low
temperature for some particular choices of parameters. In other
three cases such behavior only takes place at the extreme parameter value
corresponding to bosons and all other parameter values yield a Fermi-like
distribution at $T=0$ with continuous but non-smooth Fermi level as a
function of $g$.  Moreover the results for the permutation $\pi_{XX}$ and
$\pi_{\bowtie}$ are identical at $T=0$.  Comparing these examples with
the two simplest permutations $\pi_{id}$ and $\pi_{X}$, one can observe
that the permutations $\pi_{XX}$, $\pi_{XI}$ and $\pi_{\bowtie}$ are
similar to the inverse one except for the fact that for these permutations
the Fermi levels are non-smooth functions of the parameter
$g$~(\ref{FermiXX}, \ref{FermiXI}).
Finally the permutation $\pi_{IX}$ is close to the identical one but
the transition between the Bose- and Fermi-like behaviors occurs at
a different value of $g$ ($g=1/2$ for the identical
permutation and $g=1/3$ for $\pi_{IX}$).

\section{Conclusion}

Summarizing our discussion, let us mention some possible physical
applications of the constructions in this paper. The simplest
example of $\pi$-statistics, obeyed by the identical permutation, has
been recently considered in~\cite{Pol} as an alternative to the
state-counting procedure for exclusion statistics. However  one can note
that similar statistics (up to constant), with integer positive values of
the statistical parameter $\al$, appeared  long ago in the theory of
statistics of donor and acceptor levels in
semiconductors~\cite{Semic}. In the same manner, the  Hubbard model with
an infinite value of  Coulomb interaction on a site can be considered as a
gas of $g$-ons obeying $g=2$  statistics (a site with two single
electron levels can be occupied by only one electron).  Apparently, there
are many other systems where such statistics arise naturally.

Another example of the $\pi$-statistics with $\pi\neq id$ can be
represented by a system of anyons on a torus in a strong magnetic field
when only the lowest Landau level is occupied. As was shown in
reference~\cite{FL}  the number of many-body states in this case is given by

$$
D = s \Phi \cdot \frac{(\Phi + N(1-\al)-1)!}{(\Phi -\al N)! N!} \ ,
$$
where $\Phi$ is the magnetic flux seen by the particle and $\al$ is the
statistical parameter which is presented as $\al = k/s$ with positive
coprime  integers $k$ and $s$.
One can see that this expression corresponds to the number of many-body
states for $\pi$-statistics with the permutation

$$
\pi_0 = 0 \ ,\qquad
\pi_l = N-l \mbox{ for } 1\leq l\leq N-1 \ ,
$$
which can be termed $\pi_{1X}$ and is illustrated in Fig.6.
Let us remember that for the same system on a sphere the number of
many-body states is described by Haldane-Wu statistics with the
permutation $\pi_{X}$. So it is natural to expect the appearance of more
complicated permutations on higher genus surfaces or taking into account
higher Landau levels. Moreover, one can imagine a lot of possible physical
speculations based on the statistics with $\pi_{XX}$ or
$\pi_{XI},\pi_{IX}$. We will return to this subject in our forthcoming
paper.

In conclusion, we
considered the generalization of Haldane's state-counting procedure to
describe all possible types of exclusion statistics which are linear in the
deformation parameter~$g$. The statistics are parametrized by elements of
the symmetric group of the particles.  For several particular
cases we derived the equations for distribution
functions which generalize results obtained by Wu. Using them we
analyzed the low-temperature behavior and thermodynamic properties of these
systems and compared our results with previous studies of
the thermodynamics of a gas of $g$-ons. We speculated on the correlation
between statistical properties of gas obeying $\pi$-statistics and the
properties of the permutation $\pi$.  Physical examples where these
constructions are realized were discussed.

\section*{Acknowledgments.}

We want to thank A.S.Stepanenko for useful discussions.
This work was partially supported (K.N.I) by the Grant of the International
Science Fundation N R4T000, Grants of Russian Fund of Fundamental
Investigations N 94-02-03712 and N 95-01-00548, Euler stipend of German
Mathematical Society, INTAS-939 and by the UK EPSRC Grant GR/J35221.

\newpage

\begin{center}
{\bf Figures}
\end{center}

\begin{center}
\begin{picture}(225,70)(0,0)
\multiput(50,10)(25,0){7}{\circle*{2}}
\multiput(50,60)(25,0){7}{\circle*{2}}
\put(50,60){\line(3,-1){150}}
\put(100,60){\line(1,-1){50}}
\put(150,60){\line(-1,-1){50}}
\put(200,60){\line(-3,-1){150}}
\end{picture}
\end{center}
\begin{center}
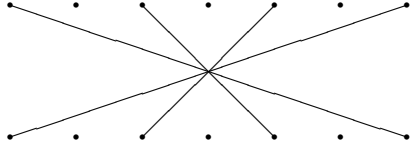
Fig.1 Illustration for the permutation $\pi_{X}$(Haldane-Wu state-counting
procedure).
\end{center}

\begin{center}
\begin{picture}(225,70)(0,0)
\multiput(25,10)(25,0){8}{\circle*{2}}
\multiput(25,60)(25,0){8}{\circle*{2}}
\put(25,60){\line(3,-2){75}}
\put(100,60){\line(-3,-2){75}}
\put(125,60){\line(3,-2){75}}
\put(200,60){\line(-3,-2){75}}
\end{picture}
\end{center}
\begin{center}
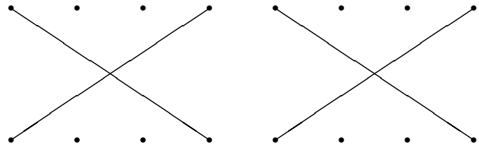
Fig.2 Illustration for the permutation $\pi_{XX}$.
\end{center}

\begin{center}
\begin{picture}(225,70)(0,0)
\multiput(25,10)(25,0){8}{\circle*{2}}
\multiput(25,60)(25,0){8}{\circle*{2}}
\put(25,60){\line(3,-2){75}}
\put(100,60){\line(-3,-2){75}}
\multiput(125,60)(25,0){4}{\line(0,-1){50}}
\end{picture}
\end{center}
\begin{center}
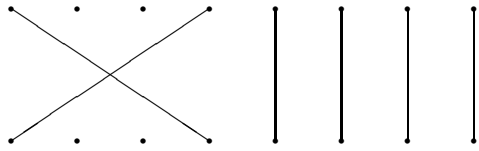
Fig.3 Illustration for the permutation $\pi_{XI}$.
\end{center}

\begin{center}
\begin{picture}(225,70)(0,0)
\multiput(25,10)(25,0){8}{\circle*{2}}
\multiput(25,60)(25,0){8}{\circle*{2}}
\put(125,60){\line(3,-2){75}}
\put(200,60){\line(-3,-2){75}}
\multiput(25,60)(25,0){4}{\line(0,-1){50}}
\end{picture}
\end{center}
\begin{center}
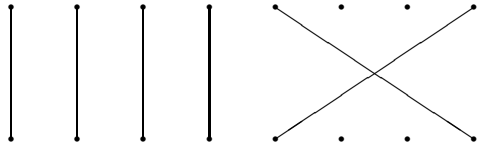
Fig.4 Illustration for the permutation $\pi_{IX}$.
\end{center}

\begin{center}
\begin{picture}(225,70)(0,0)
\multiput(25,10)(25,0){8}{\circle*{2}}
\multiput(25,60)(25,0){8}{\circle*{2}}
\multiput(25,60)(25,0){4}{\line(2,-1){100}}
\multiput(125,60)(25,0){4}{\line(-2,-1){100}}
\end{picture}
\end{center}
\begin{center}
Fig.5 Illustration for the permutation $\pi_{\bowtie}$.
\end{center}

\begin{center}
\begin{picture}(225,70)(0,0)
\multiput(25,10)(25,0){8}{\circle*{2}}
\multiput(25,60)(25,0){8}{\circle*{2}}
\put(50,60){\line(3,-1){150}}
\put(200,60){\line(-3,-1){150}}
\put(25,60){\line(0,-1){50}}
\end{picture}
\end{center}
\begin{center}
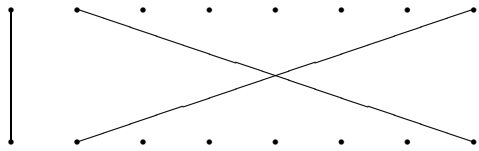
Fig.6 Illustration for the permutation $\pi_{1X}$.
\end{center}

\end{document}